\documentclass[prd,preprint,tightenlines,preprintnumbers,nofootinbib,eqsecnum,superscriptaddress]{revtex4}

\usepackage[T1]{fontenc}
\usepackage{dcolumn}
\usepackage{bm}
\usepackage{epsfig}
\usepackage{amsfonts}
\usepackage{amssymb,amscd}
\usepackage{tabularx}
\usepackage{booktabs,multirow}
 \usepackage{amsmath} 

\def\lsim{\raise0.3ex\hbox{$<$\kern-0.75em\raise-1.1ex\hbox{$\sim$}}}

\def\gsim{\raise0.3ex\hbox{$>$\kern-0.75em\raise-1.1ex\hbox{$\sim$}}}

\newcommand{\be}{\begin{equation}}

\newcommand{\ee}{\end{equation}}

\def\beq{\begin{equation}}

\def\eeq{\end{equation}}

\def\beqa{\begin{eqnarray}}

\def\eeqa{\end{eqnarray}}

\newcommand{\ba}{\begin{eqnarray}}

\newcommand{\ea}{\end{eqnarray}}

\def\gappeq{\mathrel{\rlap {\raise.5ex\hbox{$>$}}

{\lower.5ex\hbox{$\sim$}}}}

\def\lappeq{\mathrel{\rlap{\raise.5ex\hbox{$<$}}

{\lower.5ex\hbox{$\sim$}}}}

\def\Toprel#1\over#2{\mathrel{\mathop{#2}\limits^{#1}}}

\begin{document}

\title{Neutrino trident process at a muon collider}

\author{Reinaldo {\sc Francener}}
\email{reinaldofrancener@gmail.com}
\affiliation{Institute of Physics and Mathematics, Federal University of Pelotas (UFPel), \\
  Postal Code 354,  96010-900, Pelotas, RS, Brazil. 
  \vspace{0.5cm}}

\author{Victor P. {\sc Gon\c{c}alves}}
\email{barros@ufpel.edu.br}
\affiliation{Institute of Physics and Mathematics, Federal University of Pelotas (UFPel), \\
  Postal Code 354,  96010-900, Pelotas, RS, Brazil. 
  \vspace{0.5cm}}

\begin{abstract}
\vspace{0.3cm}
A TeV-scale muon collider enables the production of a highly collimated high-energy neutrino flux by muon decay in the straight sections of the collider ring. Recent studies have demonstrated that these neutrinos can be used for high-precision measurements in neutrino physics with a dedicated far-forward neutrino detector. In this study, we investigate the potentiality of the muon collider to discover  the rare neutrino trident process, which is characterized by the fusion between a weak boson from the neutrino and a photon from the Coulomb field of the nucleus target. Our results indicate large neutrino trident rates per year [$\mathcal{O}(10^3)$] in a baseline 10 kg neutrino detector for a configuration of different flavour pairs of charged leptons in the final state, which will allow  high-precision tests of the Standard Model and searches for New Physics. As a first example, we investigate the impact of the  a new gauge boson $Z'$, as predicted by the $L_\mu - L_\tau$ model, and derive the projected sensitivity to this new gauge boson. Our findings show that a neutrino detector at a muon collider will significantly extend the current excluded regions of the parameter space.
\end{abstract}


\maketitle

\vspace{1cm}

\section{Introduction}

The precision tests of the Standard Model (SM) and the search for new physics will require a new generation of particle colliders able to reach a multi-TeV center of mass energy in the next decades \cite{EuropeanStrategyGroup:2020pow}. An alternative is the  circular Muon Collider (MuCol) characterized by muon beams with an energy of 5 TeV \cite{Accettura:2023ked,InternationalMuonCollider:2024jyv,InternationalMuonCollider:2025sys}, and that has gained attention recently with the formation of the International Muon Collider Collaboration, motivating several works outlining the physics opportunities that could be reached with $\mu^+ \mu^-$ collisions \cite{Accettura:2023ked,InternationalMuonCollider:2024jyv,InternationalMuonCollider:2025sys,AlAli:2021let,Black:2022cth}. One of the main challenges of a high-luminosity MuCol is the short lifetime of muons, which would decay continuously through the collider ring and degrade the beam. However, such large decay rates of muons generate a high-intensity neutrino flux with energies ranging from sub-GeV to few TeV in straight sections of the MuCol experiment, which can be used to improve our understanding about neutrino physics (See, e.g., Refs.~\cite{Marzocca:2025inb,Adhikary:2024tvl,Burk:2026fox,Choi:2026yzw}).

The production of large neutrino fluxes in colliders has already been explored at the Large Hadron Collider (LHC) with FASER \cite{Feng:2017uoz,FASER:2018ceo,FASER:2018bac,FASER:2018eoc} and SND@LHC \cite{SHiP:2020sos,SNDLHC:2022ihg} experiments, which have observed the most energetic neutrinos made by humans so far \cite{SNDLHC:2023pun,FASER:2023zcr,FASER:2024hoe,FASER:2024ref}. These neutrino experiments were approved to continue operations during the High Luminosity (HL) LHC era. Additional neutrino experiments were proposed to upgrade the LHC neutrino program at the Forward Physics Facility (FPF) \cite{Feng:2022inv,FPF:2025bor}, which is also expected to operate during the HL-LHC period. Neutrinos at the LHC are produced exclusively in the decay of particles produced in hadron-hadron collisions, with the associated flux being characterized by the contribution of all neutrino flavors and an energy spectrum determined by the modeling of the light and heavy meson production at very forward rapidities in hadronic collisions. Such  correlation between the energy spectrum and 
the  description of soft and hard processes in hadronic collisions is currently the main source of uncertainty for the incoming neutrino flux at FASER and SND@LHC experiments. In contrast, a neutrino detector at the MuCol, denoted hereafter as the MuCol$\nu$ experiment,  would benefit from the fact that the neutrino flux from the muon beams will have a pure flavor composition and well-defined energy spectrum,    in addition to a sizable flux from particle production and propagation \cite{Burk:2026fox,Choi:2026yzw}. Such a perspective motivates one to investigate the possibility of probing rare neutrino processes in muon colliders.

\begin{figure}[t]
	\centering
	\begin{tabular}{cc}
\includegraphics[width=0.48\textwidth]{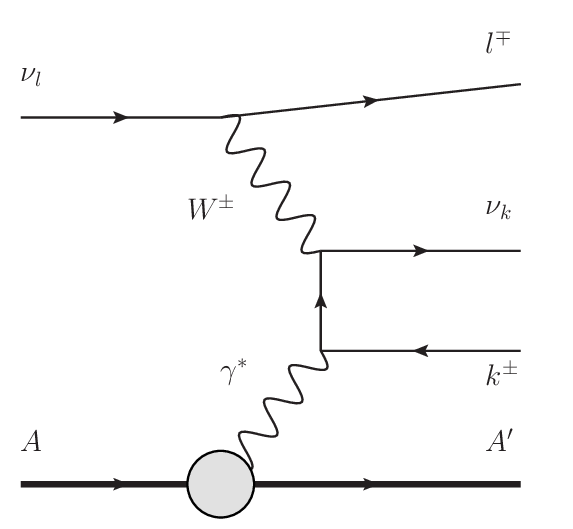} &
\includegraphics[width=0.48\textwidth]{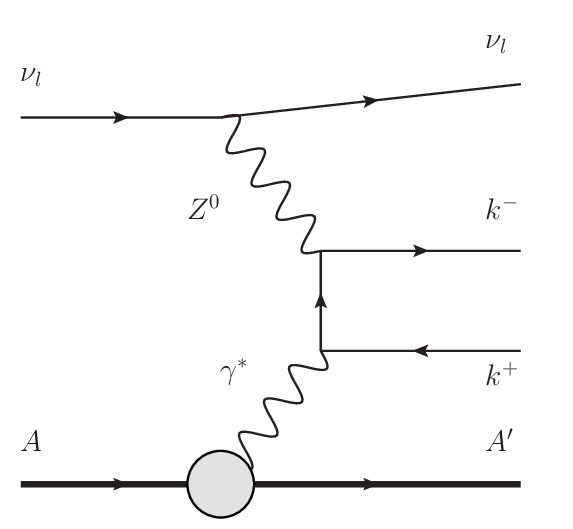} 
	\end{tabular}
\caption{ Feynman diagrams for the neutrino trident scattering off a nucleus target $A$ associated with a $W^\pm$ (left) and $Z^0$ (right) exchange. The lepton flavors $l$ and $k$ can be equal or different. }
\label{fig:diagrams}
\end{figure}

In this paper, we perform a study of the neutrino trident process at the MuCol$\nu$ experiment. Such a weak process is represented by the Feynman diagrams of Fig. \ref{fig:diagrams} and is characterized by the  production of a pair of leptons through the interaction between a weak boson, $W^\pm$ or $Z^0$, emitted by the  neutrino generated in the muon decay, and the Coulombian field of the target nucleus. The scattering can be coherent, when the nucleus interacts as a whole particle ($A = A'$), or incoherent, when the photon is exchanged by individual nucleons inside the nucleus ($A \neq A'$). The observation of this process was claimed by CHARM-II \cite{CHARM-II:1990dvf} and CCFR \cite{CCFR:1991lpl} experiments, which have observed $55 \pm 16$ and $37 \pm 12$ candidates, respectively, with pairs of muons in the final state. These observations are in agreement with the SM expectations, but its large uncertainties do not allow claiming the discovery of this reaction. This process was also searched with NuTeV \cite{NuTeV:1999wlw} and NOvA \cite{Bowles:2025mqf} experiments, but the small number of events allowed only a superior limit for its cross-section. Recent studies indicated that the observation of the neutrino trident process is feasible at the HL-LHC with FASER$\nu$(2) and SND@LHC \cite{Francener:2024wul,Altmannshofer:2024hqd,Francener:2024jra,Francener:2026ova} as well as at neutrino facilities \cite{Ballett:2018uuc,Altmannshofer:2019zhy,Bigaran:2025vea}. Our main goal in this work is to present for the first time the event rates of the neutrino trident process at the MuCol$\nu$ experiment considering the contribution of coherent and incoherent scatterings for the production of different flavor configurations in the final state. As we will demonstrate below, large neutrino trident rates are expected in this detector, which will allow us to improve our understanding of the Standard Model (SM) and searching for signals of New Physics. In particular, we will demonstrate that future measurements of the neutrino trident process at MuCol$\nu$ will be able to constrain the axial and vector SM couplings $g_V$ and $g_A$ with unprecedented precision. In addition, we will investigate the possibility of searching for leptophilic mediators using the neutrino trident process. In our analysis, we will consider the $L_\mu - L_\tau$ model \cite{He:1990pn,He:1991qd}, which implies the existence of a  new massive and neutral gauge boson, denoted  $Z'$, that couples only with leptons from the second and third generations, and will estimate the impact of the associated diagrams on the predictions for the neutrino trident cross-section. Our results indicate that uncovered excluded regions of coupling and $Z'$ mass can be reached at  MuCol$\nu$.

This paper is organized as follows. In the next section, we present a brief description of the characteristics assumed for the  MuCol$\nu$ experiment and of the theoretical framework used to estimate the neutrino trident cross-section. In section \ref{sec:res}, our predictions for the  number of neutrino trident scattering events are presented. In particular, we present the predictions for the coherent and incoherent contributions considering different combinations of lepton flavors in the final state. The impact of typical kinematical cuts is also discussed. In addition, the sensitivities to modifications in the axial and vector couplings and to the presence of a $Z^{\prime}$ are investigated. Finally, in section \ref{sec:sum}, we summarize our main results and conclusions.

\section{Calculation methodology}
\label{sec:for}

In this section, we will present the ingredients needed to estimate the number of trident neutrino events at the MuCol$\nu$ experiment. One has that such a quantity is given by
\begin{equation}
    N_\mathrm{events} = \int_{0}^{E_\mathrm{max}} \mathrm{d}E_\nu\, N_\nu \mathcal{P}_\nu (E_\nu)\, \sigma_{\nu A} (E_\nu) \frac{\rho L}{m_A} \, ,
    \label{eq:events}
\end{equation}
where $\sigma_{\nu A} (E_\nu)$ is the trident neutrino-nucleus cross-section for a neutrino energy $E_{\nu}$, $N_{\nu}$ is the rate of incoming neutrinos, and $\rho$, $m_A$ and $L$ are the material density, mass of the nucleus, and length of the detector, respectively. In particular, for MuCol$\nu$ with 10 kg of tungsten ($\rho = 19.3$ g/cm$^3$) and a radius of 10 cm, we have $L = 1.65$ cm.
Moreover, $\mathcal{P}_\nu (E_\nu)$ is the neutrino spectrum, which is determined by the muon decay rate in a muon collider, and $E_\mathrm{max} = (1+\beta) / 2E_\mu$ is the maximal neutrino energy, with  $\beta = \sqrt{1 - m_\mu^2 / E_\mu^2}$ being the muon velocity. 

Initially, let's discuss the expected characteristics of the future muon collider, which determines $N_{\nu}$. The proposed muon collider consists of a chain of accelerators and a 10 km long storage ring, with beams of $2 \cdot 10^{12}$ muons with 5 TeV injected every 0.2 seconds \cite{InternationalMuonCollider:2025sys}. With these characteristics, $10^{10}$ muons will decay in each 10 m of the ring per second. The main source of collimated neutrinos would be in the 500 m straight section of the ring present in the current design, where it is planned to be the interaction point and the main detector of the MuCol. Neutrinos produced in muon decays have transverse momentum smaller than half of the muon mass, producing a very collimated beam with an angular distribution with an angular spread of $\langle \theta \rangle \lesssim 0.1$ mrad. An additional source of spread of the neutrino flux produced comes from the spread of the muon beam, which is expected to be on the order of 0.6 mrad \cite{InternationalMuonCollider:2025sys}. These small deviations of the neutrino flux allow a neutrino detector with 10 cm of radius located at 500 m downstream of the interaction point to coverage about 10\% of the neutrinos produced in the 500 m of the straight section of the muon ring \cite{Kling:2025zsb}. Therefore, in our results, we will consider a neutrino flux with $ N_\nu = 5\cdot 10^{10}$ muonic neutrinos per second, or $N_\nu = 5\cdot 10^{17}$ muonic neutrinos per year, crossing the transverse area of the neutrino detector, in addition to the same quantity of electronic antineutrinos.

The second ingredient needed to estimate $ N_\mathrm{events}$ is the neutrino spectrum  $\mathcal{P}_\nu (E_\nu)$. Distinctly from the neutrino flux generated in hadronic collisions, which is currently impacted by the existing theoretical uncertainty present in the description of the hadron production at very forward rapidities, the neutrino spectrum associated with the muon decay can be precisely estimated. In particular, the muon neutrino and antineutrino electron spectra associated with the 
 $\mu^{-} \rightarrow e^- \nu_\mu \bar{\nu}_e$ reaction are given at the tree level by \cite{Marzocca:2025inb}
\begin{equation}
\begin{aligned}
    \mathcal{P}_{\nu_\mu} (E_\nu) =& \frac{3}{E_{\mathrm{max}}} \left[ 5-9 \left( \frac{E_\nu}{E_{\mathrm{max}}}\right)^2 + 4 \left( \frac{E_\nu}{E_{\mathrm{max}}}\right)^3 \right]  \\
    \mathcal{P}_{\bar{\nu_e}} (E_\nu) =& \frac{2}{E_{\mathrm{max}}} \left[ 1-3 \left( \frac{E_\nu}{E_{\mathrm{max}}}\right)^2 + 2 \left( \frac{E_\nu}{E_{\mathrm{max}}}\right)^3 \right] \, ,
\label{eq:spectra}
\end{aligned}
\end{equation}
where the electron mass has been disregarded.

The quantities $\rho$, $m_A$ and $L$ are depend on the characteristics of the neutrino detector. Currently, the design of a forward neutrino detector for a muon collider is still an open question. As in previous studies~\cite{Marzocca:2025inb,Adhikary:2024tvl}, we will assume a design based on that proposed for the spectrometer and calorimeter systems of FASER2 at the FPF. In particular,  we will assume a conservative detector with 10 kg operating during one year, with tungsten being the material target for neutrino interactions. Event numbers can be straightforwardly rescaled for larger detectors.

Finally, the  neutrino-nucleus cross-section for the neutrino trident process will be estimated using  the Monte Carlo (MC) generator described in Ref.~\cite{Altmannshofer:2019zhy}, which has implemented the full $2\rightarrow 4$ kinematics of the process, without assuming the validity of the equivalent photon approximation. Such a generator was adapted for a tungsten target in Ref.~\cite{Francener:2024wul}, with the nucleus being described by the Woods-Saxon nuclear charge distribution \cite{Woods:1954zz} with parameters from \cite{DeVries:1987atn}. 
In particular, we will estimate the contribution of coherent and incoherent elastic interactions. In these interactions,  the leptonic system scatters
on the full nucleus or with the individual nucleons inside the
nucleus, respectively, with final states being distinct. While in a coherent interaction the nucleus remains intact, in
the incoherent one, it is expected to break up, generating 
hadronic activity in addition to the leptonic system, which is
associated with the fragments of the nucleus. Such distinct
topology can be used, in principle, to separate the coherent and incoherent events.

\begin{figure}[t]
	\centering
	\begin{tabular}{ccccc}
\includegraphics[width=0.33\textwidth]{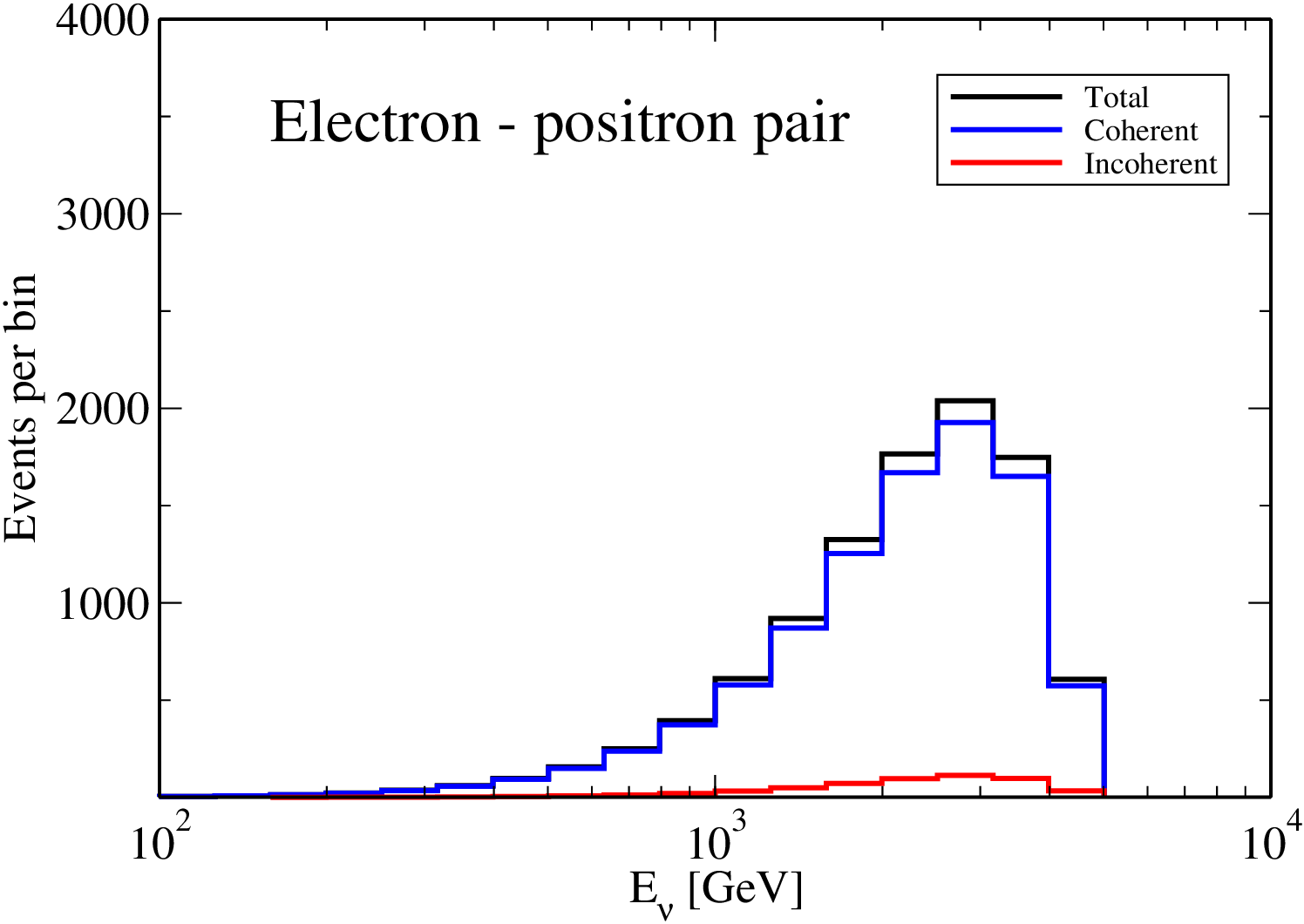} &
\includegraphics[width=0.33\textwidth]{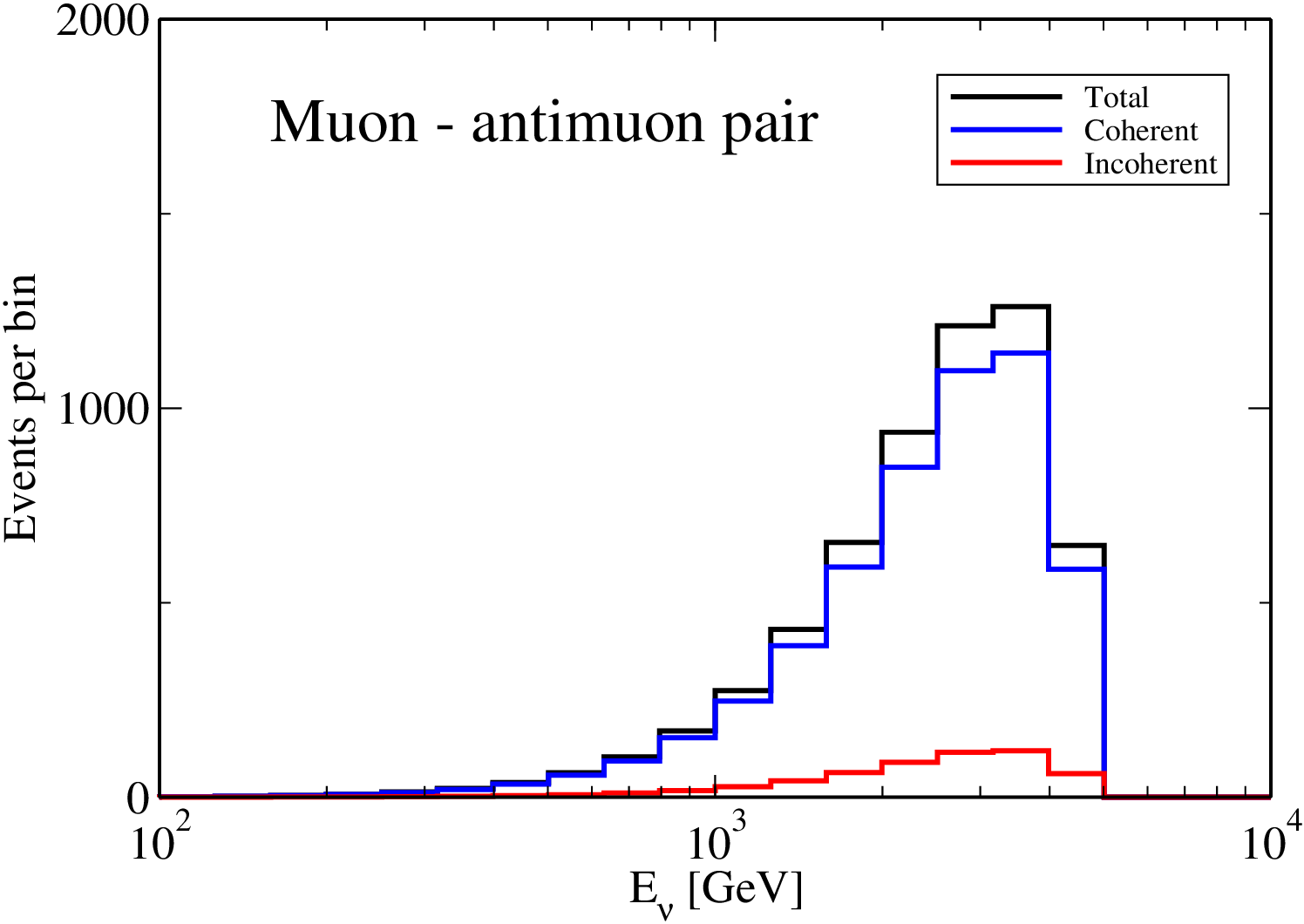} & \includegraphics[width=0.33\textwidth]{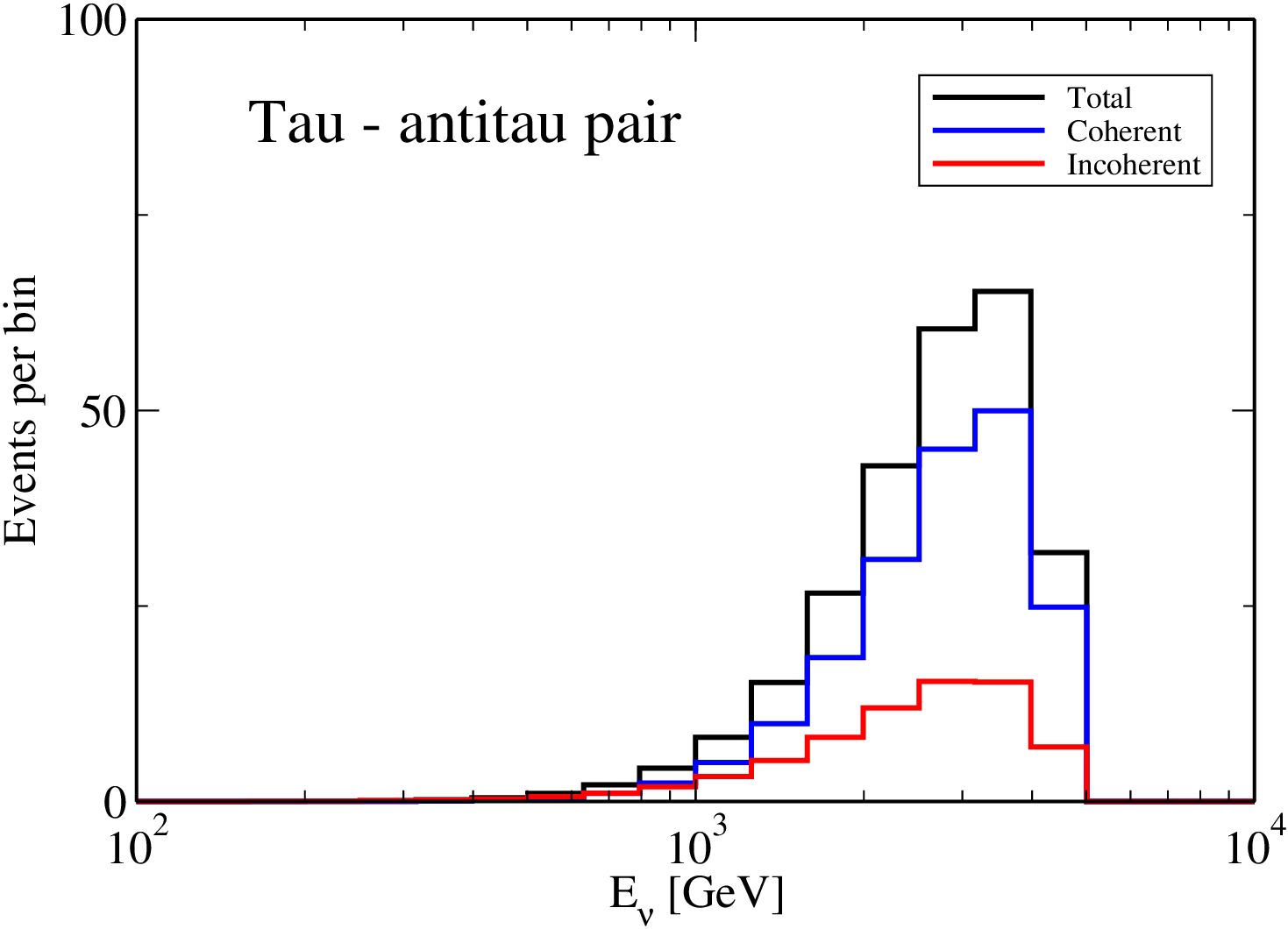} \\
\includegraphics[width=0.33\textwidth]{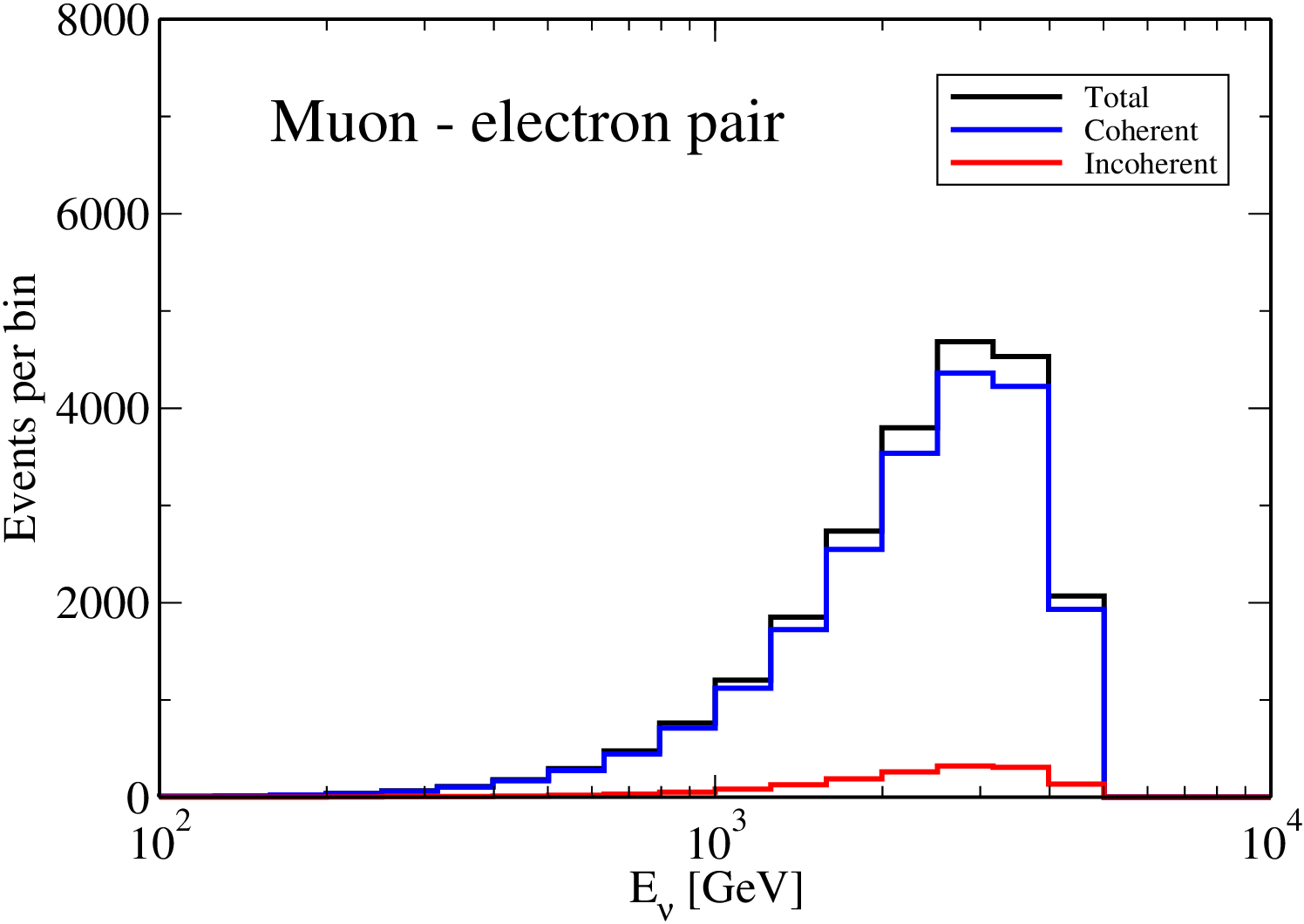} &
\includegraphics[width=0.33\textwidth]{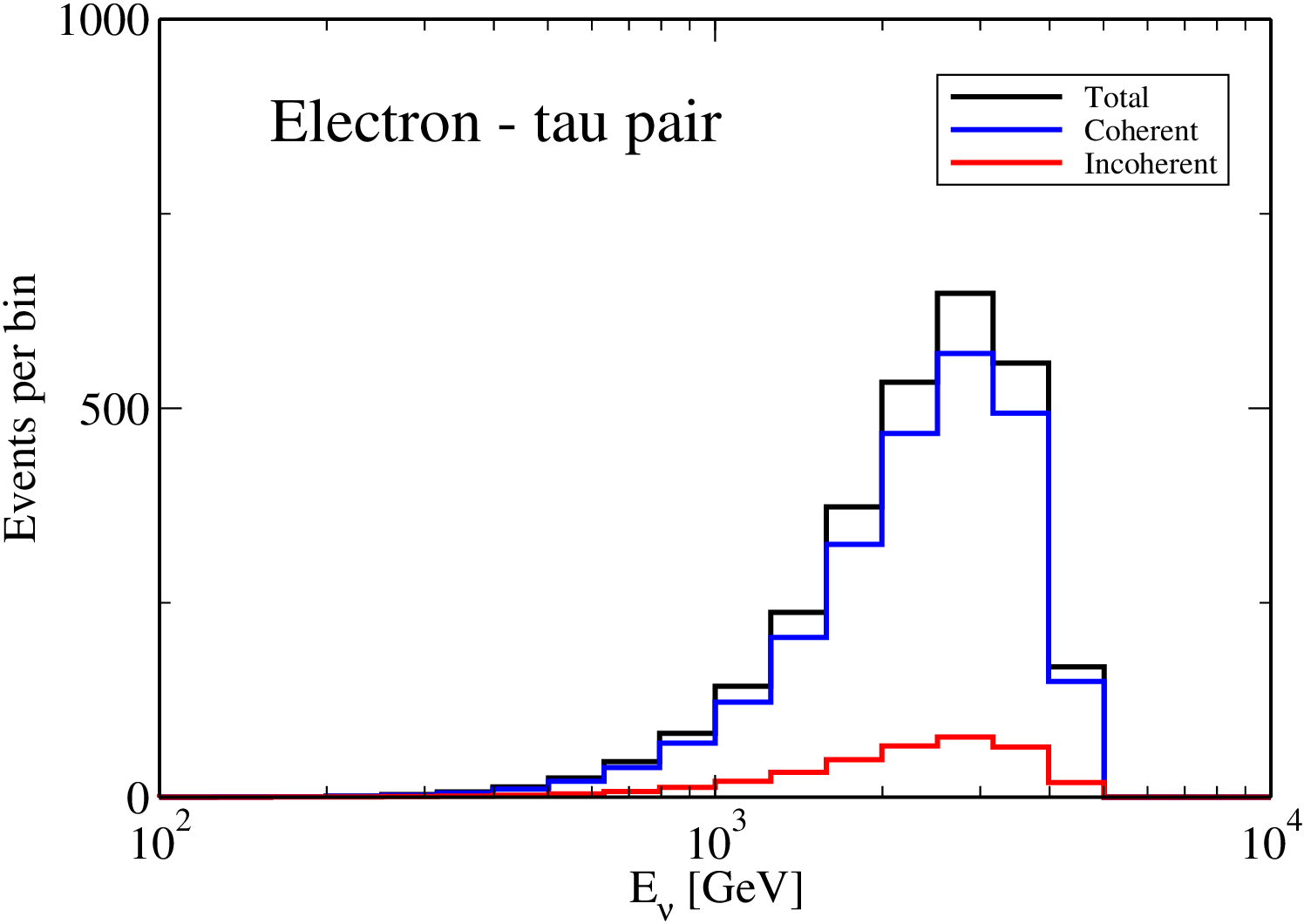} & \includegraphics[width=0.33\textwidth]{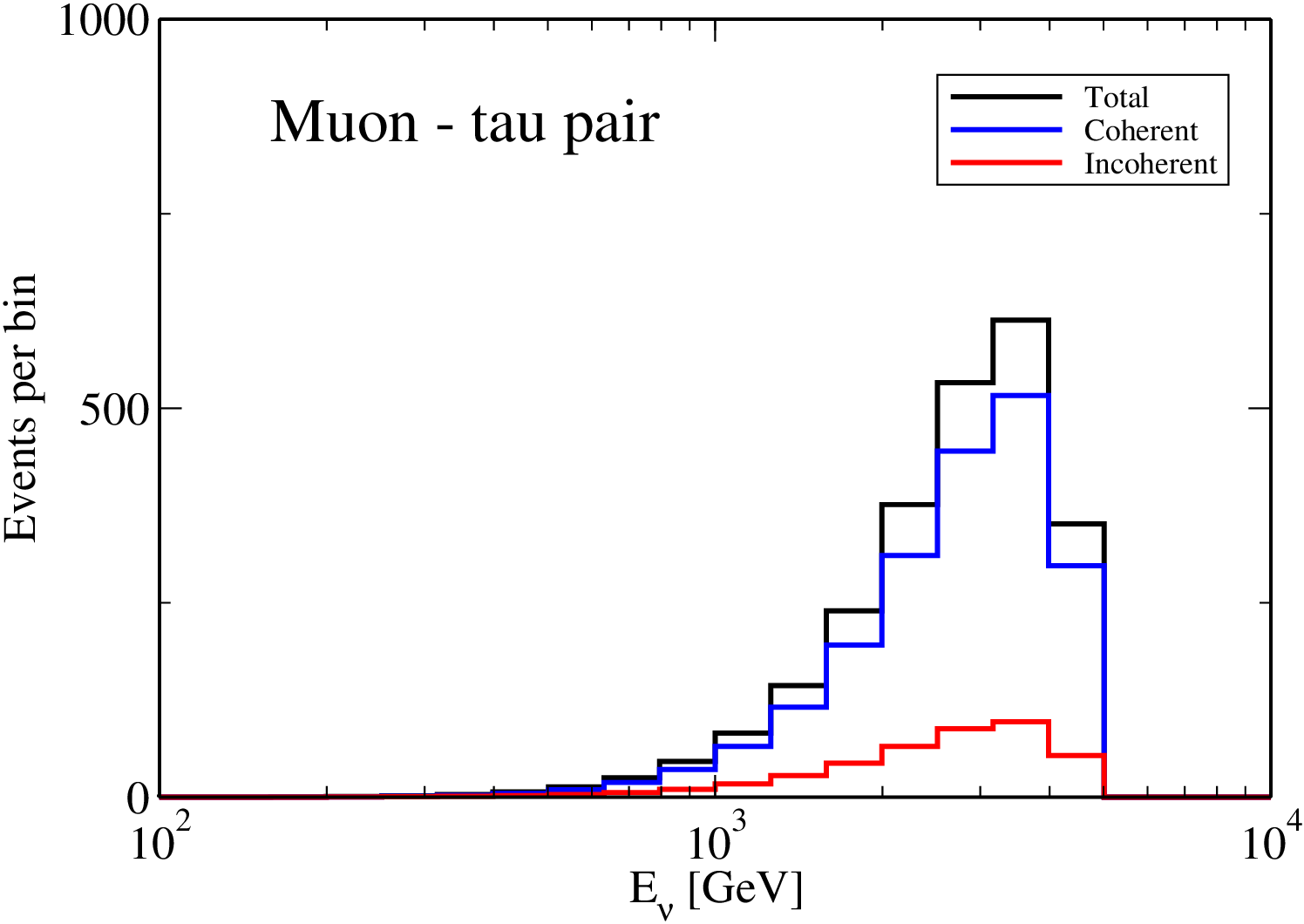} 
	\end{tabular}
\caption{ Predictions for the number of neutrino trident scattering events binned in the energy of the incident neutrino in a 10 kg neutrino detector at the muon collider during 1 year. The results for pairs of charged leptons of the same (different) flavors are presented in the upper (bottom) panels. }
\label{fig:events}
\end{figure}

\section{Results}
\label{sec:res}
Initially, in Fig. \ref{fig:events} we present our results for the number of events associated with the neutrino trident process at the MuCol$\nu$ experiment binned in the incident neutrino energy. Our predictions are for a 10 kg detector of tungsten operating during 1 year. We present results for events with a pair of charged leptons of the same family (top panels) and for charged leptons of different flavors (bottom panels) in the final state. In addition, we present predictions for coherent interactions (blue lines), incoherent interactions (red lines), and their sum (black lines). Our results indicate a peak in the event number for neutrinos with few TeV, given the energy dependence of the cross-section. We also have that this peak is process dependent, with pairs of heavier leptons in the final state with the peak at higher energies of the incident neutrino. Fig. \ref{fig:events} also shows that coherent events dominate over incoherent ones. Such result is expected, since coherent interactions are proportional do $Z^2$, while the incoherent process is linearly dependent of the atomic number. The incoherent contribution has larger importance for heavier leptons in the final state, which can be understood by the fact that more energetic photons are needed to produce them, and these  are more abundant in incoherent scatterings.
 
In Table \ref{table:Nevents_MuCol} we show the  number of neutrino trident events per year at the MuCol$\nu$ experiment associated with the incoherent and coherent interactions, as well for the sum of them, assuming distinct pairs of charged leptons in the final state. Our results indicate that the expected number of pairs of electrons/muons/taus is of the order  of $10^4$/$5\cdot 10^3$/$200$. In contrast, for pairs with different flavors, we predict values between $2\cdot 10^3$ and $20\cdot 10^3$  events per year. In general, lighter lepton states have larger event numbers, except for pairs of distinct families, which have only the contribution of the left diagram of Fig. \ref{fig:diagrams}. In this case,  the destructive interference between $W^\pm$ and $Z^0$ diagrams does not contribute.

\begin{table}[t]
\centering
\renewcommand{\arraystretch}{1.5}
\begin{tabularx}{\textwidth}{XXXl}
\toprule
\hline
\multicolumn{4}{c}{\bf Events for neutrino trident scattering at MuCol$\nu$}\\
\hline
\midrule
{\bf Final state}      & {\bf Incoherent} & {\bf Coherent}      & {\bf Total} \\
\hline	
\toprule   
$e^{+} + e^-$          & $ 549.15$        & $ 9.52 \times 10^3$ & $10.07 \times 10^3$ \\
\hline
\midrule
$\mu^{+} + \mu^-$      & $ 561.67$        & $ 5.29 \times 10^3$ & $ 5.85 \times 10^3$ \\
\hline
\midrule
$\tau^{+} + \tau^-$    & $  70.56$        & $ 0.19 \times 10^3$ & $ 0.26 \times 10^3$ \\ 
\hline   
\midrule
$e^{\pm} + \mu^\mp$    & $1563.26$        & $21.30 \times 10^3$ & $22.87 \times 10^3$ \\ 
\hline
\midrule
$e^{+} + \tau^-$       & $ 357.42$        & $ 2.48 \times 10^3$ & $ 2.84 \times 10^3$ \\  
\hline
\midrule
$\mu^{-} + \tau^+$     & $ 417.14$        & $ 2.02 \times 10^3$ & $ 2.44 \times 10^3$ \\ 
\hline
\bottomrule
\end{tabularx}
\vspace{0.3cm}
\caption{ Predictions for the number of neutrino trident events per year at the MuCol$\nu$ assuming a detector of 10 kg. We are summing over muonic neutrino and electronic antineutrino contributions. Results for coherent and incoherent interactions and distinct leptonic final states.
}
\label{table:Nevents_MuCol}
\end{table}

The previous results have been estimated without considering any experimental cuts or detector efficiencies. In what follows, we will include some of the possible experimental cuts that can be present in a future study of the trident neutrino process at the MuCol$\nu$ experiment. In particular, we will follow the procedure proposed in Ref.~\cite{Altmannshofer:2024hqd}, where the authors have identified the main backgrounds and outlined a strategy to identify signal events of the trident scattering in neutrino experiments at the LHC, and we will assume a detector with similar capabilities. We will consider the following experimental cuts in our results: ($i$) an energy efficiency $<1$ in the reconstruction of muons with less than 30 GeV, as  given in Ref.~\cite{Altmannshofer:2024hqd}; ($ii$) an angular separation between the two muons smaller than 0.1 rad; and ($iii$) in the case of incoherent interactions with a proton target, we will select only events with protons having less than 0.3 GeV of momentum, to characterize the absence of hadronic activity visible in the final state. It is important to emphasize that Ref.~\cite{Altmannshofer:2024hqd} has demonstrated that the background contributions associated with other processes with two muons in the final state, or pions misidentified as muons, can be reduced to negligible levels by applying the mentioned cuts in addition to requiring that both muons come from the same vertex. In Table \ref{table:Nevents_MuCol_cuts} we present our results for pairs of muons in the final state before and after experimental cuts, for coherent and incoherent interactions. Our results indicate that $\approx64\%/19\%/48\%$ of coherent/incoherent (p)/incoherent (n) events survive after cuts, indicating $\approx 3500$ events per year in a 10 kg detector. Such results indicate that, in principle, a precise measurement of the neutrino trident scattering will be feasible in the muon collider.

\begin{table}[t]
\centering
\renewcommand{\arraystretch}{1.5}
\begin{tabularx}{\textwidth}{XXl}
\toprule
\hline
\multicolumn{3}{c}{\bf Events with $\mu^+ \mu^-$ final state at MuCol$\nu$}\\
\hline
\midrule
{\bf Process}    & {\bf Before cuts} & {\bf After cuts}      \\
\hline	
\toprule   
Coherent (W)     & $5285.6$          & $3384.7$ \\
\hline
\midrule
Incoherent (p)   & $450.1$           & $86.5$ \\
\hline
\midrule
Incoherent (n)   & $111.6$           & $53.4$ \\ 
\hline   
\midrule
Total            & $5847.3$          & $3524.6$ \\ 
\hline
\bottomrule
\end{tabularx}
\vspace{0.3cm}
\caption{ Predictions for the number of neutrino trident scattering with pair of muons in the final state at the MuCol$\nu$ per year assuming a detector of 10 kg. Results for event number before and after experimental cuts.
}
\label{table:Nevents_MuCol_cuts}
\end{table}

\begin{figure}[t]
	\centering
	\begin{tabular}{ccccc}
\includegraphics[width=0.6\textwidth]{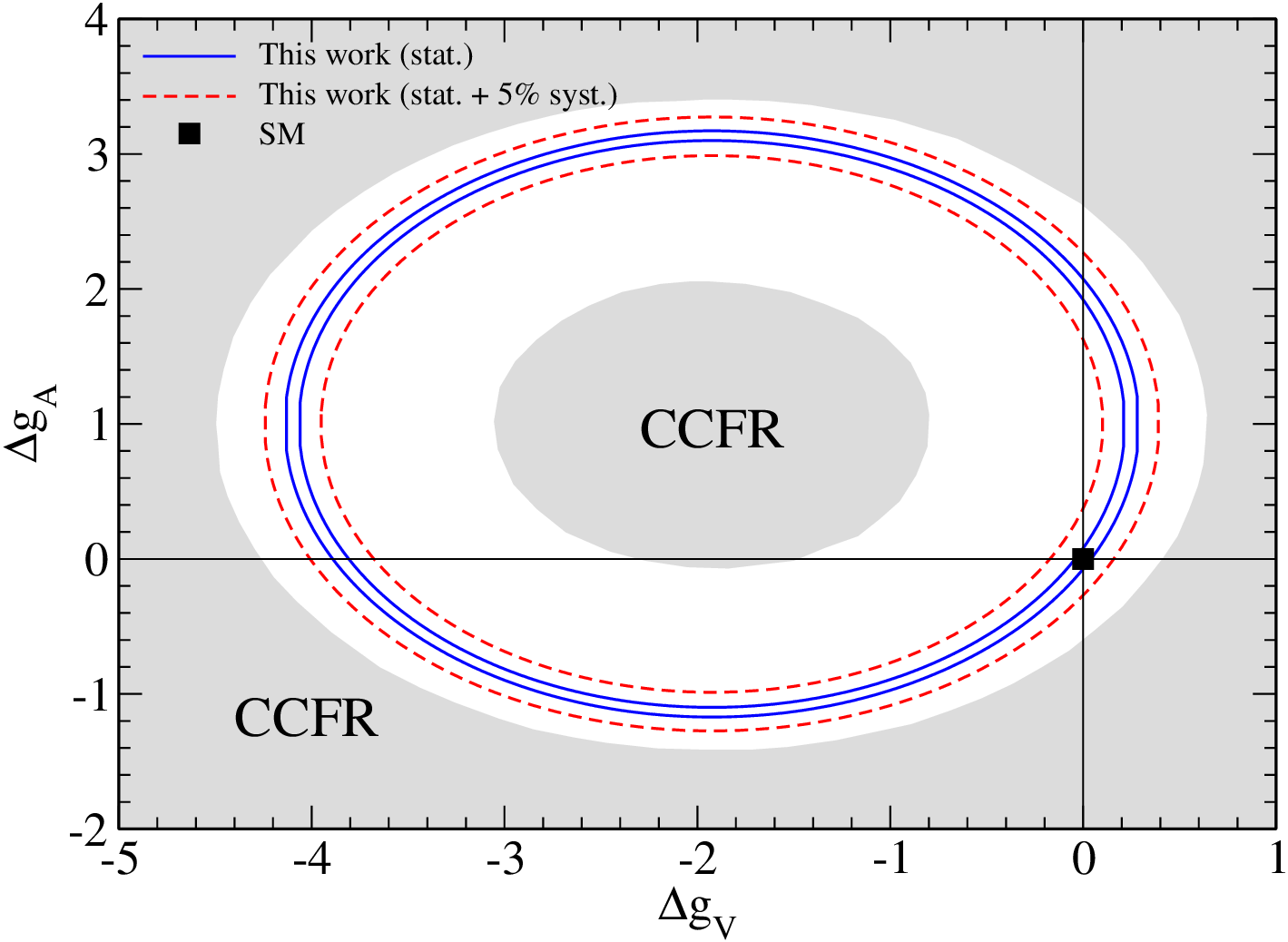} 
	\end{tabular}
\caption{ Sensitivity of the MuCol$\nu$ for the neutrino trident events associated with modifications in axial and vector couplings. Results, at the 95\% confidence level, derived for pairs of muons in the final state after experimental cuts. We have considered two scenarios: only statistical uncertainty (continuous blue line) and also assuming a systematic uncertainty of 5\% (dashed red line). For comparison, the region excluded by CCFR experiment is also presented \cite{Altmannshofer:2019zhy}. }
\label{fig:gVgAsensitivity}
\end{figure}

Motivated by these large event rates  for the trident neutrino process, in what follows we will investigate the potentiality of this process for searching for signals of beyond the standard model physics. If the new physics is heavy and suppressed by an energy scale $\Lambda$, it can be described by the Standard Model Effective Field Theory (SMEFT)\cite{Buchmuller:1985jz}, where four-lepton contact interactions appear from dimension-six operators in the Warsaw basis.
Focusing on the neutrino trident scattering induced by a muonic neutrino with pairs of muons in the final state, the new contributions come from the following term in the effective Lagrangian 
\begin{equation}
    \mathcal{L}_\mathrm{eff} \supset \frac{C^{ll}_{\mu\mu\mu\mu}}{\Lambda^2} (\bar{l}_\mu \gamma^{\mu}P_Ll_\mu)(\bar{l}_\mu \gamma^{\mu}P_Ll_\mu) \, ,
\end{equation}
where $C^{ll}_{\mu\mu\mu\mu}$ is the associated Wilson coefficient. As the neutrino trident scattering at TeV energies in the SM has a typical momentum transfer that is considerably smaller than the electroweak scale, constraints in the Wilson coefficient $C^{ll}_{\mu\mu\mu\mu}$ can be mapped onto specific deviations of the SM vector and axial-vector couplings constants $g_V^{\mu\mu\mu\mu}$ and $g_A^{\mu\mu\mu\mu}$, respectively. These couplings have the form
\begin{equation}
    \begin{aligned}
        g_V^{\mu\mu\mu\mu} \equiv & g_V = 1 + 4\,\mathrm{sin}^2\,\theta_W + \Delta g_V \\
        g_A^{\mu\mu\mu\mu} \equiv & g_A = -1 + \Delta g_A \, ,
    \end{aligned}
\end{equation}
where $\theta_W$ is the Weinberg angle. 
In Fig. \ref{fig:gVgAsensitivity}, we present our predictions for the sensitivity of the neutrino trident scattering with pairs of muons in the final state to modifications in $g_V$ and $g_A$ at 95\% confidence level (CL). For comparison, the current excluded region by the CCFR experiment, derived in \cite{Altmannshofer:2019zhy}, also at 95\% CL, is presented in gray. The SM expectation consists of $\Delta g_V = \Delta g_A = 0$. The current allowed region corresponds to the white ring. The predictions are presented considering a 10 kg detector collecting data during one year with only statistical uncertainties (continuous blue line) and assuming a 5\% systematic uncertainty as well (dashed red lines). Our results indicate that the current bounds can be significantly improved. In particular, the region inside the inner ellipse can be excluded by a decrease in the cross-section due to the interference term, while the external region  of the outer ellipse is excluded by an increase in the cross-section.

\begin{figure}[t]
	\centering
	\begin{tabular}{ccccc}
\includegraphics[width=0.33\textwidth]{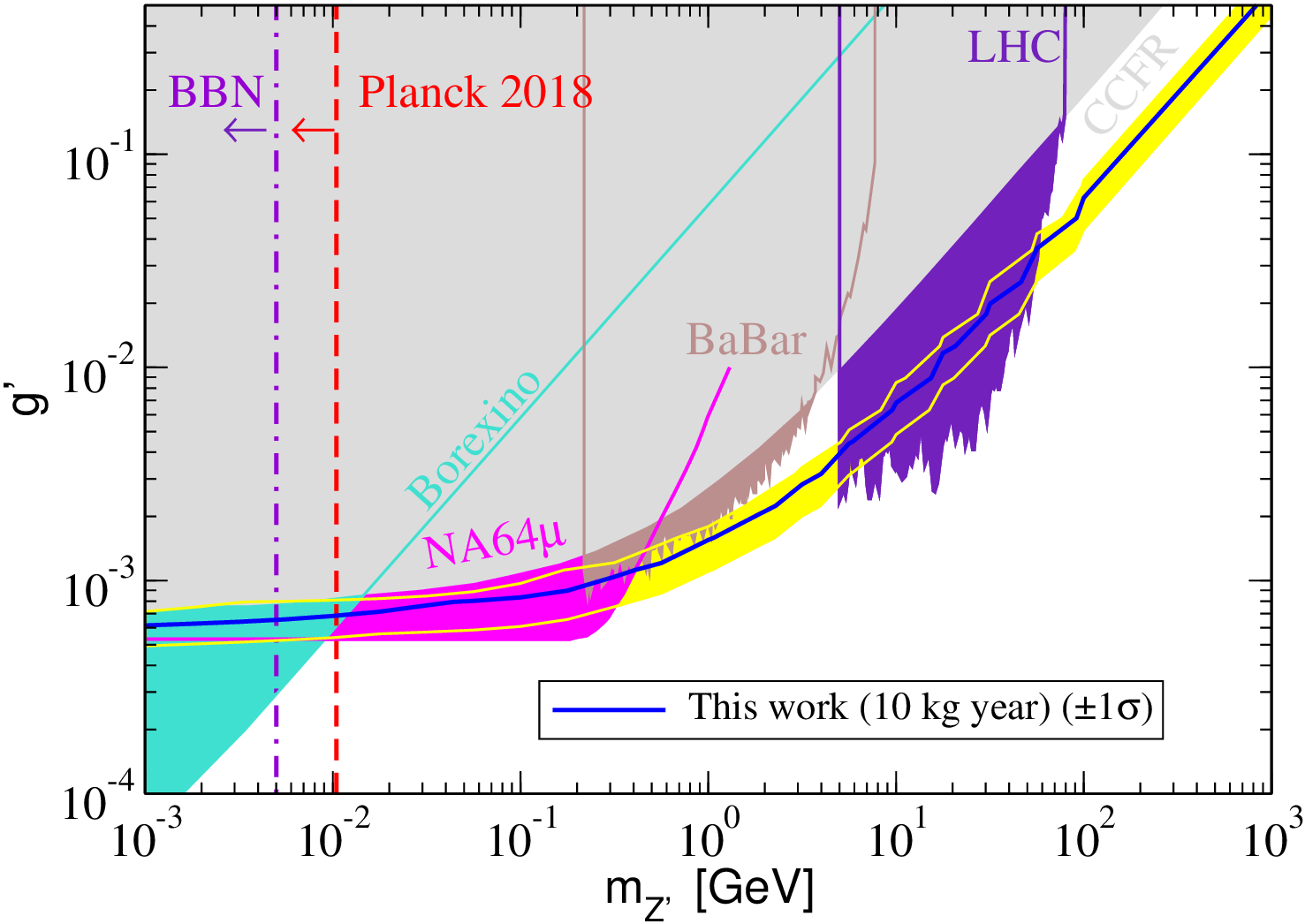} &
\includegraphics[width=0.33\textwidth]{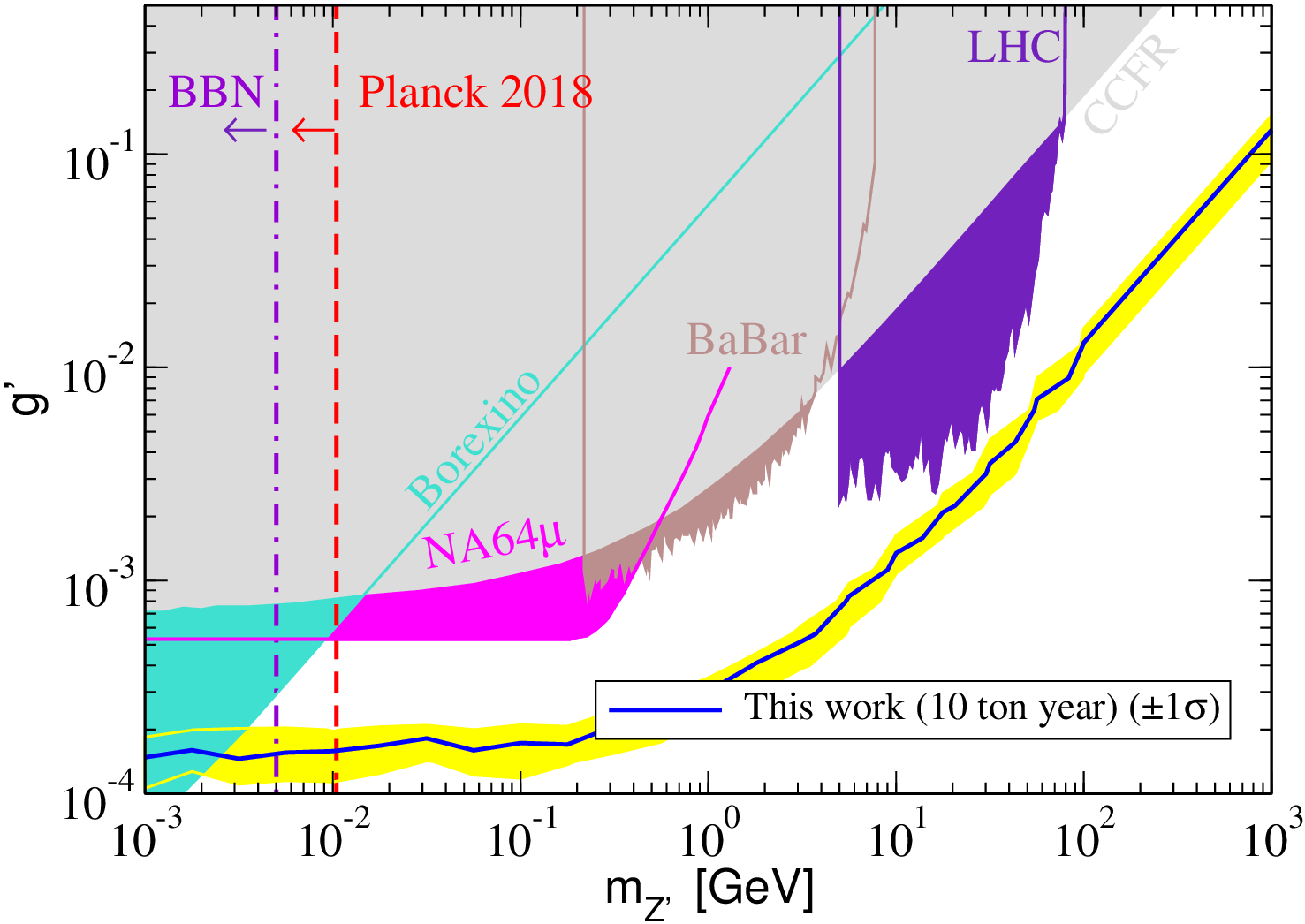} & \includegraphics[width=0.33\textwidth]{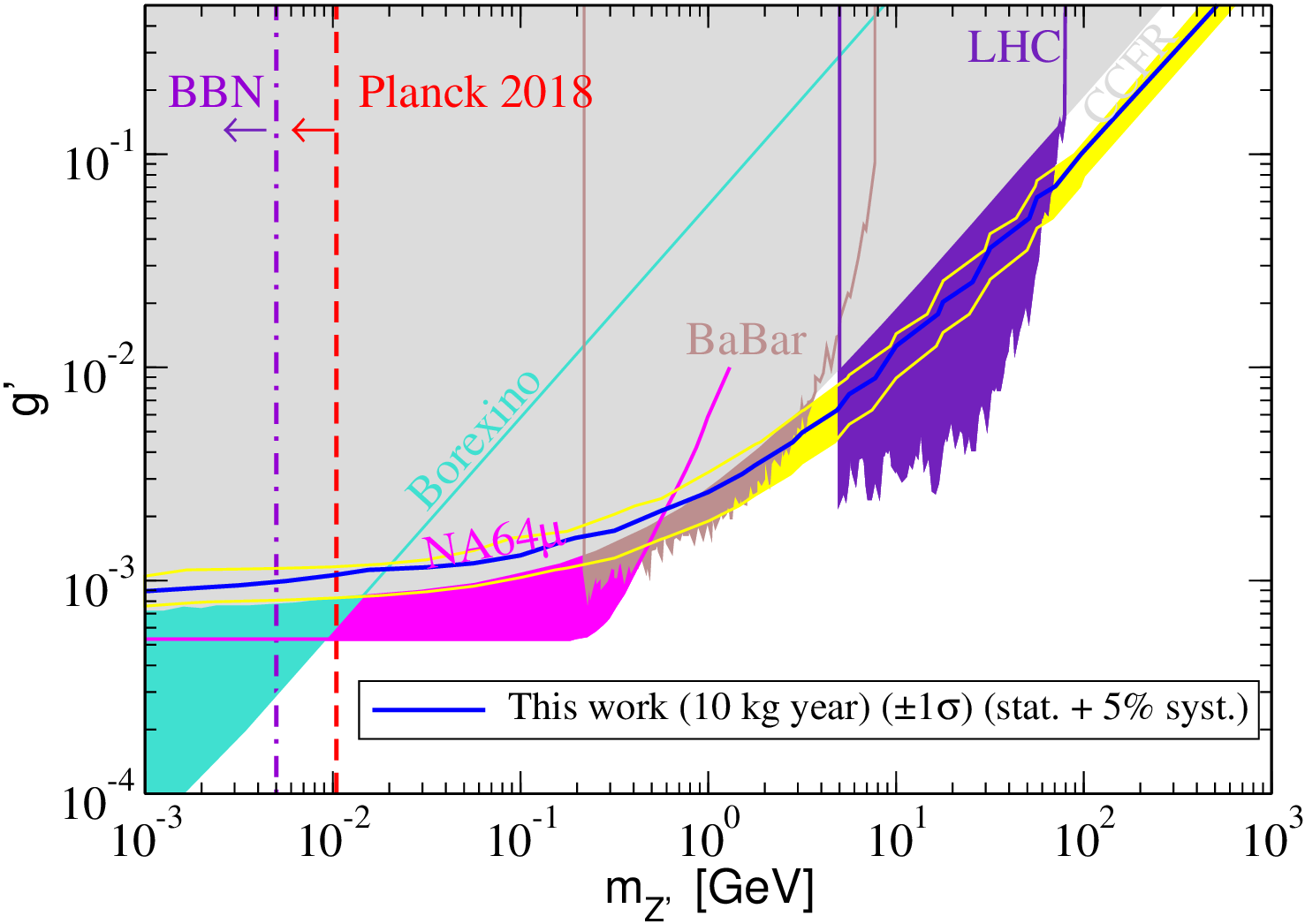} 
	\end{tabular}
\caption{ Sensitivity of the MuCol$\nu$ for the neutrino trident events associated with the $Z'$ gauge boson predicted by the $L_\mu - L_\tau$ theory. Results, at the 95\% confidence level, derived for pairs of muons in the final state after experimental cuts. We have considered three scenarios: only statistical uncertainty and a 10 kg detector (left panel); only statistical uncertainty and a 10 ton detector (middle panel); and a 10 kg detector with statistical plus a systematic uncertainty of 5\% (right panel).
For comparison, the  existing constraints from other processes and experiments are also presented. The LHC band represents the region excluded in $pp$ collisions at the LHC, considering the analysis performed by the CMS \cite{CMS:2018yxg} and ATLAS \cite{ATLAS:2023vxg,ATLAS:2024uvu} collaborations of the production of $Z^{0}$ and $W^{\pm}$ bosons and subsequent decay in the $Z^{0} \rightarrow Z' \mu^{\pm} \mu^{\mp} \rightarrow \mu^{\pm} \mu^{\mp} \mu^{\pm} \mu^{\mp}$ and  $W^{\pm} \rightarrow Z' \mu^{\pm} \nu_{\mu} \rightarrow \mu^{\pm} \mu^{\mp} \mu^{\pm} \nu_{\mu}$ channels. The BaBar band represents the associated excluded region in the study of the  $Z^{0}$ boson decay process \cite{BaBar:2016sci}. We also have in the figure the excluded regions by the Borexino experiment in interactions of neutrinos with electrons \cite{Bellini:2011rx,Harnik:2012ni}, which was constructed considering that the interaction rate should not be more than $8\%$ greater than the SM prediction. {The magenta band represents the recent results from the NA64$\mu$ experiment with 90\% CL \cite{NA64:2024klw}}. The bounds from Big Bang Nucleosynthesis (BBN) derived in \cite{Escudero:2019gzq} and from Planck 2018 \cite{Planck:2018vyg,Ghosh:2024cxi}, which implies that $m_{Z'} \gtrsim 5 \,\mathrm{MeV}$ and $m_{Z'} \gtrsim 10 \,\mathrm{MeV}$, respectively, are also presented. Finally, the excluded region derived by the CCFR experiment in the neutrino trident process with two muons in the final state \cite{CCFR:1991lpl,Altmannshofer:2014pba} is also presented. }
\label{fig:Z'sensitivity}
\end{figure}

As a second example, let's consider  the $L_\mu - L_\tau$ model \cite{He:1990pn,He:1991qd}, which gauges the difference between lepton numbers from muon and tau families, and predicts the existence of a new massive and neutral gauge boson, $Z'$. This new mediator can couple with leptons from second and third generations, with the associated Lagrangian being given by
\begin{equation}
    \mathcal{L}_\mathrm{int} = g' Z_{\alpha}' [
    (\bar{\mu} \gamma^{\alpha} \mu) - 
    (\bar{\tau} \gamma^{\alpha} \tau) + 
    (\bar{\nu}_{\mu} \gamma^{\alpha} P_{L} \nu_{\mu}) - 
    (\bar{\nu}_{\tau} \gamma^{\alpha} P_{L} \nu_{\tau})
    ]  \, ,
    \label{eq:lagrangianZ'}
\end{equation}
where $g'$ is the gauge coupling and $P_L$ is the chirality projector.  The impact of this new gauge boson can be evaluated by performing the following replacement 
\begin{equation}
    g_{V} \rightarrow g_{V} - \frac{\sqrt{2}}{2G_F} \frac{g'^2}{q_{Z'}^2 - m_{Z'}^2} \, ,
    \label{eq:Z'modification}
\end{equation}
with $m_{Z'}$ being the $Z'$ mass and $q_{Z'}^2$ its squared four momentum. When $m_{Z'}^2 \gg q_{Z'}^2$, the new boson can be integrated out, and the process is described by a simple modification in $g_V$.
In Fig.~\ref{fig:Z'sensitivity} we present our results, at the 95\% confidence level ($\pm1\sigma$), for the sensitivity of the MuCol$\nu$ experiment to the presence of the new gauge boson $Z'$ considering the trident process with two muons in the final state. For comparison, we also present the existing excluded regions for the mass and coupling parameter space ($m_{Z'}, g'$)  derived using the experimental data from BaBar \cite{BaBar:2016sci},  LHC \cite{CMS:2018yxg,ATLAS:2023vxg,ATLAS:2024uvu}, CCFR \cite{CCFR:1991lpl,Altmannshofer:2014pba}, Borexino \cite{Bellini:2011rx,Harnik:2012ni}, NA64$\mu$ \cite{NA64:2024klw} and Planck \cite{Planck:2018vyg,Ghosh:2024cxi} experiments. In  the left (middle) panel of Fig.~\ref{fig:Z'sensitivity}, we present the projected sensitivity considering a 10 kg (10 ton) detector operating during one year, derived assuming only statistical uncertainties. On the other hand, in the right panel we present the results for a 10 kg detector derived by adding  5\% of systematic uncertainty.  One has that for the $m_{Z'} \gtrsim 5$ GeV region, $g'$ is proportional to $m_{Z'}$. This happens because in this region, the mass of $Z'$ becomes much greater than the square four-momentum transferred by the neutrino, and the propagator becomes proportional to $1/m_{Z'}^{2}$. Conversely, for the region of $m_{Z'} \lesssim 10^{-2}$ GeV there is a saturation, because in this case the mass of $Z'$ becomes negligible compared to the square four-momentum transferred by the incident neutrino. For all three possible scenarios analyzed,  new regions in the parameter space  will be covered by a future study of the trident process at a muon collider, both in the $Z'$ region of lower masses ($\approx 4$ GeV), and for masses greater than that of the electroweak gauge bosons ($W^{\pm}$ and $Z^{0}$). In particular, considering a 10 ton detector, it will be able to probe the mass range greater than $10^{-2}$ GeV and $g'$ close to $10^{-4}$, which is not covered by current experiments.

\section{Summary}
\label{sec:sum}
Over the last years, several studies have demonstrated that a comprehensive physics program can be performed in a multi-TeV muon collider, which will improve our understanding of the Standard Model and search for BSM physics in unexplored limits. In particular, the 
installation of a dedicated forward neutrino detector will expand the physics potential of this collider, allowing us to study rare neutrino processes. In this paper,  we have investigated the potentiality of the MuCol$\nu$ experiment in the muon collider to search for the neutrino trident scattering. Our predictions for the total number of events in a baseline 10 kg detector operating during one year
 indicate that, in principle, the neutrino trident process can be discovered in such a detector. We have estimated this process with pairs of muons in the final state after experimental cuts, and found that more than 3500 events can be observed per year, indicating that it can be measured with unprecedented precision. We also analyzed the impact of such a measurement on two possible scenarios for the BSM physics. We have determined that the MuCol$\nu$ experiment will significantly extend the current excluded regions of these BSM scenarios. In summary, our results indicate that the measurement of the trident process at a muon collision is a promising way to test the Standard Model and search for signals of new physics.

\begin{acknowledgments}
\noindent The work of R.F. was supported by FAPERGS. VPG was partially financed by the Brazilian funding agencies CNPq, FAPERGS and INCT-FNA  (Process No. 408419/2024-5). 
\end{acknowledgments}

\end{document}